\documentclass[notitlepage,aps,pra,twocolumn,groupaddress]{revtex4-1}

\usepackage{setspace,avant,tabularx,mathrsfs,mathpazo}
\usepackage[utf8]{inputenc}
\usepackage[table]{xcolor}
\usepackage{colortbl}
\usepackage{indentfirst}
\usepackage{stmaryrd}
\usepackage{amssymb,amsmath,amsthm,amsbsy,amsfonts}%
\usepackage{bm,bibunits,color,chngcntr,epsfig,epstopdf,graphicx,dsfont}
\usepackage{hyperref,lipsum,makecell,mathrsfs,rotating}
\usepackage[english]{babel}
\usepackage[normalem]{ulem}

\newcommand{\be}{\begin{equation}}
\newcommand{\ee}{\end{equation}}
\newcommand{\bea}{\begin{eqnarray}}
\newcommand{\eea}{\end{eqnarray}}
\newcommand{\ket}{\rangle}
\newcommand{\bra}{\langle}

\newcommand{\I}{\mathds{1}}
\newcommand{\ra}{\rightarrow}

\def\C#1{\mathcal #1}

\definecolor{gray}{gray}{0.9}

\begin{document}
\newtheorem{theorem}{Theorem}
\newtheorem{prop}[theorem]{Proposition}
\newtheorem{corollary}[theorem]{Corollary}
\newtheorem{open problem}[theorem]{Open Problem}
\newtheorem{conjecture}[theorem]{Conjecture}
\newtheorem{definition}{Definition}
\newtheorem{remark}{Remark}
\newtheorem{example}{Example}
\newtheorem{task}{Task}

\title{A family of quantum von Neumann architecture}
\author{Dong-Sheng Wang}\thanks{wds@itp.ac.cn}
\affiliation{CAS Key Laboratory of Theoretical Physics, Institute of Theoretical Physics,
Chinese Academy of Sciences, Beijing 100190, China}
\date{\today}



\begin{spacing}{1.2}

\begin{abstract}
In this work, we develop universal quantum computing models that form a 
family of quantum von Neumann architecture,
with modular units of memory, control, CPU, internet, besides input and output. 
This family contains three generations characterized by dynamical quantum resource theory,
and it also circumvents no-go theorems on quantum programming and control.
Besides universality, such a family satisfies other desirable engineering requirements on 
system and algorithm designs, such as the modularity and programmability,
hence serves as a unique approach to build universal quantum computers.
\end{abstract}

\maketitle


A modern computer is a sophisticated physical system
consisting of modular units of memory, control, CPU, internet, 
and also input and output devices. 
Such a information processing system is often known as von Neumann architecture, 
and compared with other models for computing
such as the circuit model and Turing machine,
it is especially suitable to realize modularity, programmability, and automation,
besides universality.
It also greatly benefits algorithm designs, for instance,
a stored algorithm can be taken as input for a high-level algorithm
to design a new one out of it.

To establish quantum von Neumann architecture is central 
for quantum information science,
with quantum CPU, relying mainly on the circuit model, and quantum communication
having been successful in the past few decades~\cite{NC00}.
Constructing quantum memory~\cite{NC97} and control units~\cite{AFC14}
were more challenging,  
and recent progresses nevertheless show that these can be resolved~\cite{GST20,VC21,YRC20,W20_choi,W22_qvn}.
In particular, a prototype of quantum von Neumann architecture (QvN)
was proposed very recently~\cite{W20_choi,W22_qvn}. 
Meanwhile, an attempt to systematically unify various 
universal quantum computing models was carried out~\cite{W21_model,W23_ur},
and the resource-theoretic framework~\cite{W23_ur} proves to be useful to classify them.
In this work, we establish three models of QvN forming a \emph{family},  
each with distinguished features that applies to different practical settings.

\begin{figure}[b!]
    \centering
    \includegraphics[width=0.2\textwidth]{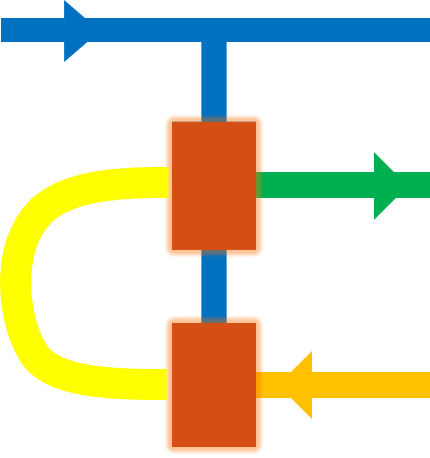}
    \caption{Schematics of the functionality of quantum von Neumann architecture.
    They are the Program, Control, CPU, Input and Output units, 
    with colors of yellow, blue, red, orange, and green, respectively.
    Besides, quantum internet is also needed to communicate quantum data.}
    \label{fig:qvn}
\end{figure}

The primary function of a QvN, shown in Figure~\ref{fig:qvn}, 
can be better understood if we compare with the classical case.
The two fundamental quantum principles are the channel-state duality~\cite{Cho75}
and Heisenberg's uncertainty principle.
The former motivates the use of the dual state of a channel, known as Choi state,
as stored quantum programs, and also the use of ebits for communication~\cite{BBC+93}, 
while the later characterizes quantum coherence~\cite{SAP17} 
which plays various roles for all components of a QvN.
A throughout description of QvN would be beyond this work,
yet the basic picture is that, with the interplay with a quantum control unit, 
quantum programs can be read out from and written into a quantum memory unit,
and can also be uploaded and downloaded with the usage of a quantum internet. 
Quantum measurements play roles for the input, output,
and also during the computation. 

A more rigorous characterization of computing models 
is based on quantum resource theory~\cite{CG19},
which is able to capture constraints such as those on space and time.
A hierarchy of universal resources was employed to define a family of  
computing models, consisting of at least three generations~\cite{W23_ur}. 
Given the set of finite-dimensional Choi states as programs,
we establish three generations of QvN, summarized in Table~\ref{tab:triplet_evo}. 
This forms the anticipated evolution family (`$e$-family')~\cite{W23_ur}, 
and indeed, we can see a hierarchy of resources of programs
from local to nonlocal ones. 
It shall also be noted that here we mainly rely on the 
resource theory of quantum programs. 
It is also possible to define resource theories
each for the task of quantum control and communication.

\begin{table}[t!]
    \centering
    \begin{tabular}{|c|c|c|c|c|c|}\hline
         & QvN-I & & QvN-II & & QvN-III \\ \hline
    $\C F$ & EBC & $\supset$ & SEPC & $\supset$ & PROC \\ \hline
    $\C O$ & SC & $\supset$ & SLOCC  & $\supset$ & S1O \\  \hline
    $\C U$ & memory & $\prec$ & bi-memory & $\prec$ & nonlocal memory  \\  \hline
    \end{tabular}
    \caption{The family of quantum von Neumann architecture (QvN) with 
    their resource-theoretic characterization: the free set $\C F$, free operations $\C O$,
    and the set of universal resources $\C U$.
    The short-hands are: 
    EBC for entanglement-breaking channels, SC for superchannels that preserve it;
    SEPC for bipartite separable channels, SLOCC for stochastic local operations with classical communication
    that preserve it;
    PROC stands for tensor product of local channels, S1O for stochastic one-local operations that preserve it.
    The relation ``a $\prec$ b'' means the resource $b$ is more powerful than $a$. 
    }
    \label{tab:triplet_evo}
\end{table}


Quantum evolution is described by completely positive, 
trace-preserving (CPTP) maps, also known as channels~\cite{NC00}.
For a finite-dimensional Hilbert space $\C H$ and the set of states $\C D(\C H)$ with it,
any quantum channel $\C E: \C D(\C H) \ra \C D(\C H)$ can be represented as its dual state
$\omega_{\C E} := \C E \otimes \I (\omega)$,  
usually known as a Choi state~\cite{Cho75}, for $\omega:=|\omega\ket \bra \omega|$,
and $|\omega\ket:= \sum_i |ii\ket /\sqrt{d}$, $d=\text{dim}(\C H)$.
The state $|\omega\ket$ is a Bell state, also known as an ebit.
Choi states are bipartite, we will call the 1st (2nd) part as `head' (`tail').
Given a program $\omega_{\C E}$, an initial input state $\rho$ is written into the tail 
by a binary measurement $\{\sqrt{\rho^t}, \sqrt{\I-\rho^t}\}$,
and the final output in terms of an 
expectation value $\text{tr}(\C A \C E(\rho))$ for observable $\C A$ 
can be read out from the head~\cite{W20_choi}.
This leads to the universality to simulate any quantum algorithm 
in the usual circuit model whose input needs to be prepared first~\cite{W22_qvn}.

Besides the input and output by measurements,
one can also do program conversion and composition. 
The conversion of a program $\omega_{\C E}$ to another 
$\omega_{\C F}=\hat{\C S}(\omega_{\C E})$ is by 
the so-called superchannel~\cite{CDP08a},
which in general can be realized by attaching ebits
and then execute separable unitary followed by measurements~\cite{W22_qvn};
namely, $\hat{\C S}(\omega_{\C E})=\text{tr}_e \C V \otimes \C U (\omega_{\C E}\otimes \omega) $,
with $\text{tr}_e$ as the trace on the head of the ebit with projection to $|0\ket\bra 0|$ on its tail,
and the unitary $\C V$ ($\C U$) acts on the wires of heads (tails). 
Therefore, a superchannel $\hat{\C S}$ can be further stored as a Choi state $\omega_{\hat{\C S}}$,
and then acted upon by another superchannel, leading to the so-called high-order operations~\cite{CDP09}. 

The composition of two programs relies on a generalization of quantum teleportation~\cite{BBC+93}. 
From a point of view of symmetry, the standard teleportation is $Z_d\times Z_d$ symmetric,
with the qudit Pauli operators as its projective representation. 
The symmetry can be extended to $SU(d)$ by grouping the nontrivial Pauli byproduct together~\cite{W20_choi}.
This enables the deterministic composition of any unitary programs 
$\{ |\omega_{U_i} \ket \}_{i\in \C I}$ to form $|\omega_{\prod_i U_i} \ket$.

Now we study the resource-theoretic formulation of QvN.
A universal resource theory is defined by four sets:
a set of free states $\C F$, a set of free operations $\C O$,
a set of resource states $\C R$, and a set of universal resources $\C U \subset \C R$~\cite{W23_ur}. 
We will see below the generations in the QvN family rely on different forms of 
programs and operations on them. 
For simplicity, we assume classical control and focus on operations on programs.

For the 1st generation denoted as QvN-I,
we require no explicit structure of the space $\C H$, 
and no constraint (e.g., on locality or energy) for the channels.
The primary feature of programs $\omega_{\C E}$ is the entanglement contained in the ebit $\omega$.
Therefore, we identify the set of entanglement-breaking (EB) channels with the form
$\C E_\text{EB}(\rho)=\sum_i \text{tr}(F_i \rho) \sigma_i$
as the free set $\C F$, for $\{F_i\}$ as a POVM, each $\sigma_i$ a state~\cite{HSR03}. 
This means that the input data $\rho$ is replaced by $\sigma_i$,
with only the probability $\text{tr}(F_i \rho)$ depending on $\rho$.
The dual state of $\C E_\text{EB}$ is separable; 
therefore, it is natural to identify the set of free operations $\C O$
as the superchannels that preserves the Choi-state form.
For instance, the local operations on the tail need to be unital as the local state
for the tail is completely mixed. 

All channels that do not break entanglement are now resources,
and the universal set $\C U$ is the unitary programs
which have the same entanglement with the ebit $\omega$.
This is analog with the resource theory of quantum memory~\cite{RBL18,LHW18}.
For QvN-I, the composition of programs can be treated as free,
while the conversion of programs by superchannels will consume ebits. 
Furthermore, it is more important to identify the computational power of the free setting. 
We find it is at least as powerful as classical computers, namely,
it can simulate any classical algorithms, 
hence can also describe a classical von Neumann architecture, for instance.

\begin{figure}[t!]
    \centering
    \includegraphics[width=0.35\textwidth]{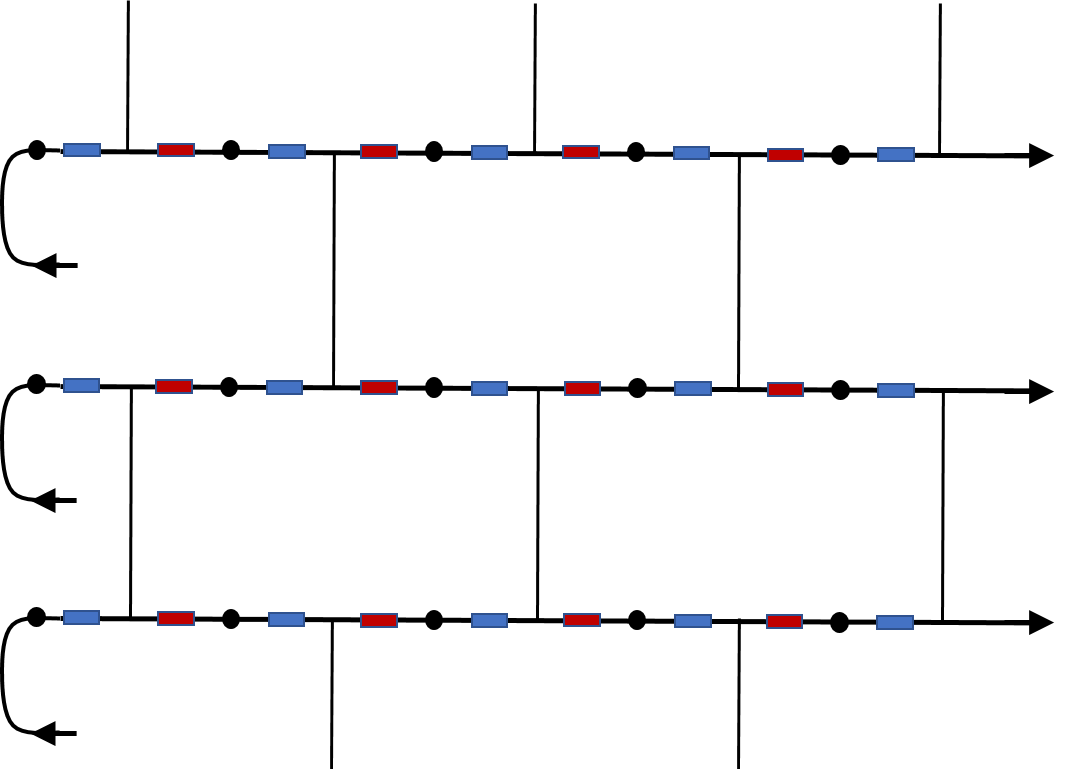}
    \caption{Schematics of a circuit in QvN-II.
    The evolution of each qubit defines a local part, 
    with the entangling gates (vertical lines) as universal resources. 
    The connection between any two programs is a composition
    via standard (blue box) or $SU(2)$ covariant (red box) teleportation.
    A black dot represents a qubit gate.
    }
    \label{fig:qvn-ii}
\end{figure}

A random or stochastic algorithm can be described by a stochastic process 
$\vec{p}\mapsto \vec{q}=S \vec{p}$ for a stochastic matrix $S$,
which can be simulated by a POVM $\{F_i\}$ with $S_{ij}=\bra j|F_i|j\ket$, 
as a special type of EB channels. 
What a classical von Neumann architecture can do with bits and pbits, e.g., 
copying, read and write can be simulated by EB channels. 
There can also be quantum coherence in the states out of EB channels;
however, the coherence cannot propagate due to the breaking of entanglement.
Whether there is an advantage over the classical case is left for 
further exploration (cf. Ref.~\cite{Hay17}). 

It is intriguing to compare the resource theory of quantum programs 
with that for coherence or entanglement, defined for the amplitude family 
based on the circuit model~\cite{W23_ur}.
In the setting of QvN-I, 
the ebit $\omega$ as a program is used to propagate quantum coherence,
while coherence itself (in a non-diagonal state~\cite{SAP17}) 
is created by measurements or unitary evolution.
So quantum program (or memory) is different from coherence as universal resources.
The ebit $\omega$ is also the universal resource of entanglement defined relative to 
biseparable states with a locality constraint.
We will see below that, actually the dynamical analog of ebit is entangling bipartite channels,
leading to the model of QvN-II.

For a bipartite system $\C H_A \otimes \C H_B$,
a bipartite channel $\C E^{AB}$ is separable if its Choi state $\omega_{\C E^{AB}}$ is separable
for the $A|B$ partition. 
It is straightforward to establish the resource theory with 
the set of separable Choi states $\omega_{\C E^{AB}}$ as the free set $\C F$,
and stochastic local operation with classical communication (SLOCC)
that preserves it as the free operation $\C O$,
and resources as entangling bipartite channels.
For instance, the controlled-not (CNOT) gate is entangling
and its entanglement is maximal for the two-qubit system.
This motivates our formulation of QvN-II.
For qubits, we identify the set of all unitary gates that are locally equivalent to CNOT as $\C U$.
The building blocks are the programs $\omega_\text{H}$, $\omega_\text{T}$, 
and $\omega_\text{CNOT}$, 
for the Hadamard gate H, T $=Z^{1/4}$ for Pauli $Z$ gate, 
forming a universal gate set $\{$H, T, CNOT$\}$~\cite{NC00}. 
The composition of them can realize any quantum algorithm.

As shown in Figure~\ref{fig:qvn-ii}, a computation in QvN-II is a tailed circuit, 
with input at the tail and output at the head, and composition in the middle. 
A composition is a teleportation acting on a tail wire and a head wire.
A qubit gate, H or T, composed before a CNOT requires a standard teleportation 
as the Pauli byproduct can pass through the CNOT gate,
while a qubit gate composed after a CNOT requires the $SU(2)$ covariant teleportation 
as in this case, the Pauli byproduct only passes through 
when the full symmetry $SU(2)$ is available.
It is clear to see all Pauli byproducts are pulled out to the end of the circuit,
hence can be corrected. 
The cost is proportional to the spacetime volume of a circuit, namely,
the total number of elementary gates in it. 

\begin{figure}[b!]
    \centering
    \includegraphics[width=0.35\textwidth]{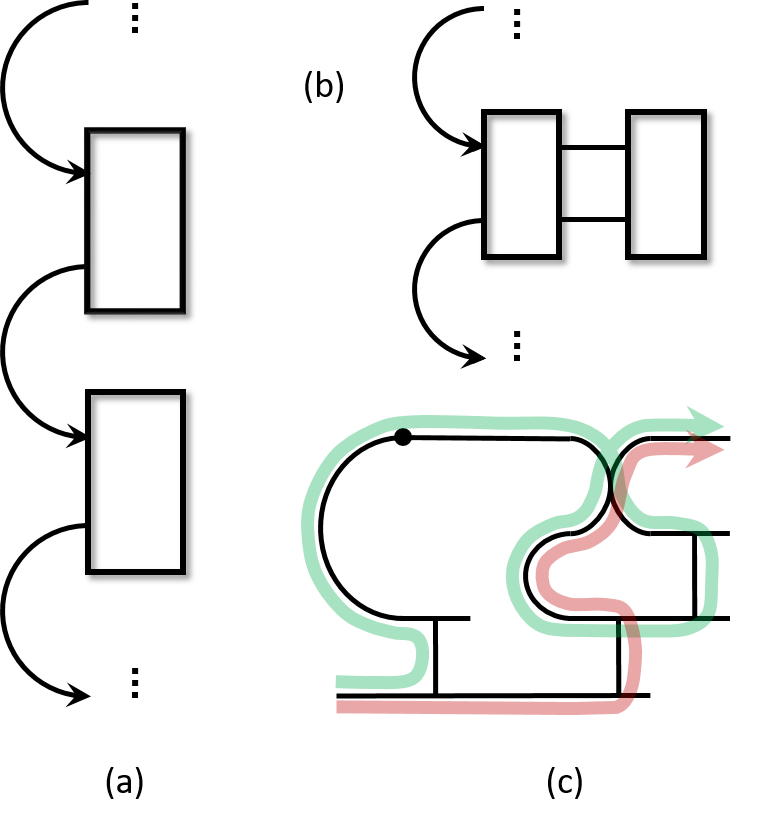}
    \caption{(a) The information flow in QvN-II. Each box is a composition. 
    (b) Schematics for the switchability. 
    (c) A switchable program with an ``on'' path (green) and ``off'' path (red). 
    Vertical lines represent CNOT gates, or CZ gates equivalently.
    The black dot is a qubit program, and the black crossing wires represent the swap gate.
    This scheme also applies to the CNOT program.
    }
    \label{fig:qvn-ii2}
\end{figure}

This model appears as a measurement-based quantum computing,
but there are key differences.
In the teleportation-based scheme~\cite{GC99},
gates are not pre-stored as their Choi states.
The CNOT gates are stored in the cluster state~\cite{RB01},
but the qubit gates are from the single-qubit measurements in rotated bases.
A more fundamental difference is that the information flow in QvN-II 
is of depth-one, see Figure~\ref{fig:qvn-ii2}(a).
Modular byproducts, the composition serves as an ebit 
which is an identity gate to propagate information. 
This leads to a key feature of tailed circuits that we name as
``switchability'' of gates (panel (b)). 
In details, an ebit is attached to a head of a program,
forming a ``triode'' of heads.
As shown in panel (c),
one needs to apply a CNOT gate between certain wires,
which is the first step in a composition.
This allows the possibility to switch on or off a gate 
by completing one of the two information flow paths
(and erasing the other by CNOTs).
Such a switchability allows the correction of wrong gates during a circuit,
and also yields novel programmability of circuits.

We now move on to construct the QvN-III model.
We shall consider multipartite or nonlocal programs
which in turn requires more restrictive free settings.
It turns out the model in Ref.~\cite{YRC20} is the right starting point.
With a generalized sine state $|\Phi\ket$ acting on $2n$ qudits,
the output $U|d\ket$ is approximated to the accuracy $\epsilon \sim \frac{1}{n^2}$
by $\int d \hat{U} (\hat{U}\otimes |\eta_{\hat{U}}\ket \bra \eta_{\hat{U}}| ) |d\ket |p_U\ket $
for a covariant POVM formed by states $|\eta_{\hat{U}}\ket$ and $d \hat{U}$ being the Haar measure
on the group $SU(d)$,
and 
a program state $|p_U\ket=(U^{\otimes n} \otimes \I) |\Phi\ket$.
Notably, this scheme also works optimally for learning or estimation,
and can also be viewed as a covariant code~\cite{YMR+22}. 
Namely, let $|f\ket=U|d\ket$ and $|f\ket \mapsto |d\ket |p_U\ket$ be the isometric encoding, 
it is then clear this encoding is $SU(d)$ covariant and the POVM realizes the decoding. 
A novel feature here is that
the covariance enables blind computing, 
which is distinct from the other two models above and motivates the following construction. 

\begin{figure}
    \centering
    \includegraphics[width=0.35\textwidth]{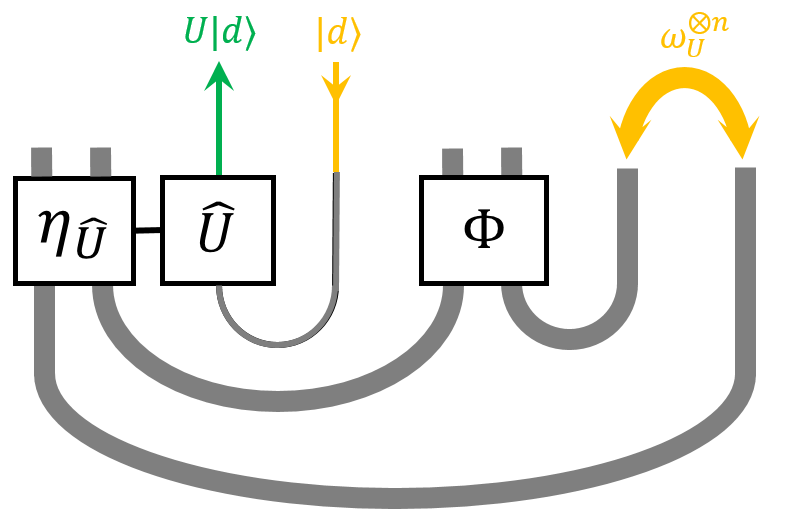}
    \caption{Schematics for the model of QvN-III.
    The bold wires are multiple ebits, 
    $\eta_{\hat{U}}$ with control to $\hat{U}$ represents the covariant POVM,
    $\Phi$ represents the circuit to prepare the generalized sine state.
    The grey part forms the program $\omega_\text{cov}$.
    Note we have rotated the diagram for convenience. 
    }
    \label{fig:qvn-iii}
\end{figure}

As shown in Figure~\ref{fig:qvn-iii},
we use a nonlocal program, $\omega_\text{cov}$, to store the whole scheme,
which can be viewed as a superchannel. 
The decoding is stored by its circuit, 
with the POVM $\{\eta_{\hat{U}}\}$ realized 
by a projective measurement according to Naimark's dilation, 
and the state $|\Phi\ket$ is stored by its preparation circuit. 
Both the program $|\omega_U\ket^{\otimes n}$ and data $|d\ket$ are injected by measurements.
Now with each qudit being a local subsystem,
it is clear that the free set $\C F$ is formed by transversal operations, 
including the read/write measurements,
the free operations $\C O$ are local and no apparent classical communication is needed,
and the program $\omega_\text{cov}$ as the universal resource $\C U$ 
enables a quasi-exact universality~\cite{WWC+22}.
It realizes $U|d\ket$ blindly if the input, output, and program are held by separate parties.
With the technique of composition,
a few blind programs $\{|p_{U_i}\ket\}$ can also be composed together.

We therefore established the family of QvN using the resource-theoretic framework,
as summarized in Table~\ref{tab:triplet_evo}.
Their universal resources are the quantum memory defined by the ebits,
bipartite memory defined by the CNOT gate, 
and the nonlocal memory from the covariant programming. 
We can also see a subset hierarchy of their free settings,
from the largest one that can efficiently simulate 
any classical algorithms, to local bipartite separable channels, 
and to one-local channels that can only generate product states. 
We should note that there can be other members in this family,
and the resource-theoretic characterization of a QvN can be more complete
by also considering resource theories for quantum control and quantum communication.


To summarize, we showed that QvN can be established using dynamical quantum resource theory. 
In the framework of QvN,   
it remains to see if QvN can benefit other studies of quantum information.
For instance,
a direct application is algorithm design~\cite{W21_model}.
Some of the recent quantum algorithms, such as 
the singular-value transformation~\cite{MRTC21},
quantum learning~\cite{HKP21} can be described as superchannels,
hence attributing their speedups properly to the resources of quantum memory. 
Prototypes of QvN can be realized on current quantum computers,
especially with the aid of high-fidelity multiple-qubit gates~\cite{LKS+19,KM20,KMN+22}.
Distinct features of QvN such as the swichablility of gates and 
the novel type of tailed circuits can be experimental investigated.


Discussions with G. Chiribella, Y.-D. Wu, Y. Yang, S. Luo, and D. Yang 
are acknowledged.
This work has been funded by
the National Natural Science Foundation of China under Grants number
12047503 and number 12105343.

\end{spacing}

\bibliography{ext}{}
\bibliographystyle{ieeetr}


\end{document}